\documentstyle[12pt,aaspp]{article}

\def\mk{\ \mu{\rm K}}
\def\cobe{{\sl COBE}}

\slugcomment{{\it Astrophysical Journal Letters}, in press}

\begin{document}
\title{Maximum-Likelihood Analysis of the COBE Angular Correlation Function}
\author{Uro\v s Seljak\altaffilmark 1 and Edmund Bertschinger}
\affil{Department of Physics, MIT 6-207, Cambridge, MA 02139 USA}
\begin{quote}
\altaffilmark{}
\altaffiltext{1}{Also Department of Physics, University of Ljubljana, Slovenia}
\end{quote}

\begin{abstract}
We have used maximum-likelihood estimation to determine the quadrupole
amplitude $Q_{\rm rms-PS}$ and the spectral index $n$ of the density
fluctuation power spectrum at recombination from the \cobe\ DMR data.
We find a strong correlation between the two parameters of the form
$Q_{\rm rms-PS}=(15.7\pm 2.6)\exp[0.46(1-n)]\mk$ for fixed $n$.  Our
result is slightly smaller than and has a smaller statistical uncertainty
than the 1992 estimate of Smoot et al.
\end{abstract}


\section{Introduction}

The search for the cosmic microwave background radiation (CMBR) anisotropies
finally yielded positive results with the detection made with the
Differential Microwave Radiometers (DMRs) on the {\it Cosmic Background
Explorer} (\cobe) satellite (Smoot et al. 1992; Bennett et al. 1992;
Wright et al. 1992).  These authors analyzed sky maps based on the first
year of \cobe\ data and presented quantitative results for the quadrupole
moment, the angular correlation function, and power spectrum parameters
characterizing the large angular scale fluctuations.  In simple cosmological
models the CMBR anisotropy is related directly to the gravitational potential
fluctuations present during recombination on large length scales
(\cite{sw67}).  The \cobe\ data thus enable quantitative testing of
cosmological theories, free from the complications of galaxy formation.

Perhaps the most important measure of CMBR anisotropy is provided by the
angular correlation function of the temperature anisotropy $\Delta T$,
$C(\alpha)=\langle \Delta T(\vec{n}_1)\Delta T(\vec{n}_2)\rangle$.
The average is taken over the sky for all pairs of directions separated by
angle $\alpha=\cos^{-1}(\vec{n_1}\cdot\vec{n_2})$.  Smoot et al. (1992,
Fig. 3) and Wright et al. (1992, Fig. 2) presented estimates for
$C(\alpha)$ with the monopole and dipole contributions removed;
Smoot et al. (but not Wright et al.) also removed the quadrupole.

Smoot et al. (1992) used a least-squares fit to the measured values
$C(\alpha_i)$ to estimate the amplitude (represented by the rms CMBR
quadrupole, $Q_{\rm rms-PS}$) and the spectral index $n$ of the density
power spectrum, obtaining $n=1.1\pm0.5$ and $Q_{\rm rms-PS}=17\pm5\mk$
for $n=1$, corresponding to the scale-invariant Peebles-Harrison-Zel'dovich
spectrum.
In order to make these estimates one needs the covariance matrix for
$C(\alpha_i)$.  There are two sources of uncertainty contributing to this
covariance matrix: measurement errors (chiefly receiver noise) and cosmic
variance (the intrinsic statistical fluctuations expected because the CMBR
temperature is measured on a surface of finite extent).  The cosmic variance,
and therefore the full covariance matrix of the measurements, depends on
the power spectrum parameters one is trying to determine.  Smoot et al.
presented least-squares estimates with and without the cosmic variance.
The estimated values changed very little but the $\chi^2$ decreased from
79 to 53 for 68 degrees of freedom when the cosmic variance was included.
We find that the inclusion of cosmic variance is even more important for
the correlation function of Wright et al. (1992, Fig. 2), as it causes
$\chi^2$ to decrease from 155 to 52 for 64 degrees of freedom if $n=1$
and $Q_{\rm rms-PS}=17\mk$.

Recently, Scaramella \& Vittorio (1993) emphasized the importance of
cosmic variance.  Including this, they reanalyzed the angular correlation
function of Smoot et al. (1992) using $\chi^2$ minimization and Monte
Carlo simulations.  They concluded that the best-fit amplitude is
$Q_{\rm rms-PS}=(14.5\pm1.7)(1\pm0.06)\mk$ for $n=1$, where the first
error bar is due to cosmic variance and the second one to measurement
uncertainties.  However, they did not simulate the actual sky sampling
and data reduction procedure applied to the data.

There are several shortcomings of the previous statistical analyses.
First, the least-squares method is inappropriate when the covariance
matrix depends on the parameters one is trying to estimate.  In this case
least-squares estimators are often biased and in general they do not have
minimum variance.  The standard least squares error estimates are also
biased, and the sum of squares of residuals does not have a $\chi^2$
distribution.  Correlations between different lag angles $\alpha_i$ are
not taken into account because the least-squares method (and also the
minimum $\chi^2$ method) uses only the diagonal part of the covariance
matrix of $C(\alpha_i)$.  As we show below, this covariance matrix has
three contributions, two of which involve cosmic variance and are
nondiagonal.  Although the biases can be corrected using Monte Carlo
simulations, least-squares estimators do not make the most efficient use
of the data because they are not minimum-variance.

Wright et al. (1993) reexamined the \cobe\ data using the rms variance
on the $10^\circ$ scale and the Boughn-Cottingham statistic (Boughn et
al. 1992).  They confirm their earlier results for the quadrupole amplitude,
concluding that the amplitude is consistent with $Q_{\rm rms-PS}=17\pm5\mk$
for $n=1$.

We have chosen instead to estimate the power spectrum parameters using
the maximum-likelihood method.  This method has the advantage of providing
estimates that are asymptotically unbiased and of minimum variance
(Eadie 1971).  This {\it Letter} presents our approximate maximum-likelihood
determination of $Q_{\rm rms-PS}$ and $n$ using realistic Monte Carlo
simulations of the \cobe\ sampling and data reduction procedures.
We will show that the uncertainties of these two parameters are strongly
correlated and that the optimal value of $Q_{\rm rms-PS}(n)$ for the
correlation functions of Smoot et al. (1992) and Wright et al. (1992)
is slightly lower and has smaller uncertainty than estimated by those authors.

\section{Statistical Method}
The measured correlation function $C(\alpha_i$) was given in $m$ ($=70$
for Smoot et al. 1992 and 65 for Wright et al. 1992) equidistant points
$\alpha_i$ between $0^{\circ}$ and $180^{\circ}$.  Approximating the
$m$-dimensional distribution as a multivariate normal distribution one
can write the likelihood function as
\begin{equation}
L(Q_{\rm rms-PS},n)=
{\exp\left [-{1\over 2}({\bf\Delta C})^T{\bf M^{-1}}({\bf\Delta C})\right ]
\over [(2\pi)^m\det({\bf M})]^{1/2}}\ .
\end{equation}
Here $({\bf\Delta C})^T$ and $({\bf\Delta C})$ are $m$-dimensional row and
column vectors with entries $(\Delta C)_i=C(\alpha_i)-\langle C(\alpha_i)
\rangle$, while ${\bf M}=\langle ({\bf\Delta C})\,({\bf\Delta C})^T\rangle$
is an $m\times m$ matrix.  The angle brackets here denote averages over
both measurement errors and the statistical ensemble of temperature
fluctuation fields for given $Q_{\rm rms-PS}$ and $n$.  Both $\langle
C(\alpha)\rangle$ and ${\bf M}$ depend on these parameters.
Maximum-likelihood estimates $\widehat Q_{\rm rms-PS}$ and $\widehat n$ are
obtained by maximizing $L(Q_{\rm rms-PS},n)$, keeping $C(\alpha_i)$ fixed
at the measured values.  The curvature of the likelihood function around
the maximum provides an estimate of the covariance matrix of errors of the
estimated values (Eadie 1971).

In assuming a multivariate normal distribution we are maximizing the wrong
likelihood function because the actual distribution of $C(\alpha_i)$
involves products of normally distributed variables (the raw temperature
measurements) and in general is not normally distributed.  However, the
assumption of a normal distribution is not crucial to the success of our
method when there are small departures from normality.  Although the
asymptotic property of the smallest variance among all estimation methods
is no longer valid, the increase in variance should be small.  One could
appeal to the central limit theorem and assume that the average over many
products of pairs is nearly normally distributed even when the individual
products themselves are not.  We have tested this possibility by making
Monte Carlo realizations of pixel maps and calculating from them the
distributions of $C(\alpha_i)$.  They are close to normal away from the
tails.  The extended tails increase the variance of the estimated values,
but the effect is small and does not significantly affect the overall
efficiency of the maximum-likelihood method.  We will use Monte Carlo
simulations with the correct distributions to estimate the bias and variance
of $\widehat Q_{\rm rms-PS}$ and $\widehat n$.

The DMR instrument measured, for two independent channels at each of
three frequencies, the differences in CMBR temperature $T_i-T_j$
between two directions $\vec n_i$ and $\vec n_j$ separated by $60^\circ$
(Smoot et al. 1990).  These differences were then fitted to give estimates
of $\Delta T_i$ (with the monopole, dipole, and possibly quadrupole
contributions removed) at 6144 points of an equal area sky map (Torres
et al. 1989; Janssen \& Gulkis 1992).  The correlation function was then
calculated by Smoot et al. (1992) and Wright et al. (1992) using
\begin{equation}
C(\alpha_k)={\sum_i N_i\sum_j N_j\ \Delta T_i\,\Delta T_j
\,\delta(\vec{n_i}\cdot\vec{n_j}-\cos\alpha_k) \over
\sum_i N_i\sum_j N_j\ \delta(\vec{n_i}\cdot \vec{n_j}-\cos\alpha_k)},
\label{calp}
\end{equation}
where $N_i$ is the number of measurements in each pixel $i$ and the
$\delta$-function indicates that the sum is performed over all different
pairs $\Delta T_i$ and $\Delta T_j$ such that $\cos^{-1}(\vec{n_i}
\cdot\vec{n_j})$ is in a given bin $k$ of $\alpha$.  The weighting by
$N_i$ gives a minimum-variance estimate for $C(\alpha_k)$.  The temperature
anisotropies $\Delta T_i$ and $\Delta T_j$ can be drawn either from the
same map (autocorrelation) or from two different maps (cross-correlation).
Smoot et al. cross-correlated the 53A+B and 90A+B GHz channels, while
Wright et al. combined cross-correlations of 53A, 53B, and 90B channels
in a manner equivalent to the autocorrelation of a single weighted map.

Evaluating the mean and covariance matrix of $C(\alpha_k)$ requires
knowing the $N_i$ and the covariance matrix of temperature measurement
errors.  These quantities depend in a complex way on the \cobe\ sky
scan pattern and on the detailed properties of the DMR instrumentation
(Boggess et al. 1992; Smoot et al. 1990).  We used a simulation program,
kindly provided by Ned Wright, to simulate the spacecraft operation for
the first year of operation, including the correct orbit, spacecraft spin,
tracking of the Sun, Moon, and planets, with data rejection if the
instrument pointed too close to the Earth and Moon, etc. (Smoot et al.
1992).  The simulated measurements were gathered into 6144 equal area
pixels using the quad-cube routines provide to us by Wright.

In addition, we calculated the covariance matrix of temperature measurement
errors for the sky maps.  The main source of measurement error is receiver
noise (Smoot et al. 1990).  Because the raw DMR measurements are temperature
differences for two beams separated by $60^\circ$, the errors in the
temperature maps (obtained by fitting to the differences) are correlated.
We estimated that the off-diagonal elements of the covariance matrix are
nearly all much smaller than 1\% of the diagonal elements (rising to 6\%
for a few elements) so that to a good approximation one can safely
neglect the noise correlations for different pixels.  One can then write
$\Delta T_i=\Delta T_i^0+\epsilon_i$, where $\Delta T_i$ is the measured
value for pixel $i$, $\Delta T_i^0$ the true value and $\epsilon_i$ the
noise contribution.  The pixel noise is normally distributed with
$\langle \epsilon_i\rangle =0$ and $\langle \epsilon_i\epsilon_j\rangle=
(\sigma^2/N_i)\delta_{ij},$ where $\sigma$ is the noise contribution
from a single measurement (Janssen \& Gulkis 1992).

The true values $\Delta T_i^0$ are also stochastic variables in theories
of large-scale structure.  The ensemble averages are
$\langle \Delta T_i^0\rangle =0$ and
$\langle \Delta T_i^0\Delta T_j^0\rangle=C^0(\alpha_k)$, where
$C^0(\alpha_k)$ is a theoretical correlation function (including
beam smearing and pixelization).  This function is most conveniently
characterized by its expansion in Legendre polynomials,
$C^0(\alpha_k)=\sum_lC_l\,G_l^2\,P_l(\cos \alpha_k)$, where $G_l$
is the window function of the DMR beam, for which we used the values
given by Wright et al. (1993) with a slight correction for beam smearing
and pixelization.  The angular power spectrum on large angular scales
for primeval adiabatic density fluctuations with $\Omega=1$ is (Bond \&
Efstathiou 1987)
\begin{equation}
C_l={(2l+1) \over 5}
\frac{\Gamma[l+(n-1)/2]\,\Gamma[(9-n)/2]}{\Gamma[l+(5-n)/2]\,
\Gamma[(3+n)/2]}\,Q^2_{\rm rms-PS}\ .
\label{be}
\end{equation}
This expression is accurate for the angular scales probed by \cobe.
To test it we replaced it with the more accurate angular correlation
function obtained with a full integration of the coupled Boltzmann and
Einstein equations for $n=1$ by Bond \& Efstathiou (1987) and found
negligible change in our estimates.

Averaging over measurement errors and an ensemble of true sky maps
we can now calculate the mean and covariance matrix of $C(\alpha_k)$
neglecting the fitting and removal of low-order multipoles.
For the cross-correlation case we get
$\langle C(\alpha_k)\rangle =C^0(\alpha_k)$; for autocorrelations
there is an additional term at $\alpha_k=0$ due to noise.  The
covariance matrix for the cross-correlation case is the sum of
three terms, ${\bf M}= \hbox{NN}+\hbox{SN}+\hbox{SS}$, which we denote as
noise-noise (NN), signal-noise (SN) and signal-signal (SS) terms.
The NN term depends only on the measurement errors, the SN term scales
as $Q^2_{\rm rms-PS}$ and the measurement variance, and SS scales as
$Q^4_{rms-PS}$.  The contributions to the matrix elements $M_{k_1\,k_2}$
are given by the following expressions:
\begin{equation}
\hbox{NN}={\sigma_A^2\sigma_B^2\,\delta_{k_1\,k_2}\over N_{\rm tot}}
\sum_{i_1}N_{i_1}\sum_{j_1}N_{j_1}\,
\delta(\vec n_{i_1}\cdot \vec n_{j_1}-\cos\alpha_{k_1})\ ,
\end{equation}
$$\displaylines{
\hbox{SN}={1\over N_{\rm tot}}\sum_{i_1}N_{i_1}\sum_{j_1}N_{j_1}\,
\delta(\vec n_{i_1}\cdot \vec n_{j_1}-\cos\alpha_{k_1})\cr
\times\biggl\{\sigma_A^2\sum_{j2}N_{j_2}\,\delta(\vec n_{i_1}\cdot
\vec n_{j_2}-\cos\alpha_{k_2})\,
C^0(\cos^{-1}[\vec n_{j_1}\cdot\vec n_{j_2}])\cr
\hfill+\sigma_B^2\sum_{i2}N_{i_2}\,\delta(\vec n_{i_2}\cdot
\vec n_{j_1}-\cos\alpha_{k_2})\,
C^0(\cos^{-1}[\vec n_{i_1}\cdot\vec n_{i_2}])\biggr\} \ ,
\hfill\llap(5)\cr}$$
$$\displaylines{
\hbox{SS}={1\over N_{\rm tot}} \sum_{i_1}N_{i_1}\sum_{j_1}N_{j_1}
\sum_{i_2}N_{i_2}\sum_{j_2}N_{j_2}\,
\delta(\vec n_{i_1}\cdot\vec n_{j_1}-\cos\alpha_{k_1})\,
\delta(\vec n_{i_2}\cdot\vec n_{j_2}-\cos\alpha_{k_2})\,\cr
\times
\biggl\{ C^0(\cos^{-1}[\vec n_{i_1}\cdot\vec n_{i_2}])\,
C^0(\cos^{-1}[\vec n_{j_1}\cdot\vec n_{j_2}])
+C^0(\cos^{-1}[\vec n_{i_1}\cdot\vec n_{j_2}])\,
C^0(\cos^{-1}[\vec n_{j_1}\cdot\vec n_{i_2}]) \biggr\}\ ,\hfill\llap(6)\cr} $$
where
$$\displaylines{N_{\rm tot}=\sum_{i_1}N_{i_1}\sum_{j_1}N_{j_1}
\sum_{i_2}N_{i_2}\sum_{j_2}N_{j_2}\,
\delta(\vec n_{i_1}\cdot\vec n_{j_1}-\cos\alpha_{k_1})\,
\delta(\vec n_{i_2}\cdot\vec n_{j_2}-\cos\alpha_{k_2})\cr}$$
and the indices range over all map pixels.  Pixels labeled with $i$
correspond to map $A$, for which the measurement variance of $\Delta T_i$
is $\sigma_A^2/N_i$, while $j$ corresponds to map $B$.  For
cross-correlations $\sigma_A\ne\sigma_B$ in general.  For autocorrelations
one sets $\sigma_A=\sigma_B$ and the NN and SN terms are increased by a
factor of 2.

We see that the NN term involves double summation over sky maps,
the SN term involves triple summation, and the SS term quadruple
summation.  Even after the galactic lattitude cuts made by Smoot
et al. (1992, $\vert b\vert>20^\circ$) and Wright et al. (1992,
$\vert b\vert>30^\circ$), one is still left with several thousand
pixels.  While the NN and SN terms can be summed exactly, the direct
calculation of SS becomes computationally too expensive. Instead, we
evaluated it using Monte Carlo simulations.  We generated 10,000 maps
for a gaussian random field $\Delta T(\vec{n})$ on the sphere
having each theoretical angular power spectrum (i.e., value of $n$)
to be tested.  For each realization the angular correlation function was
measured using equation (2) and the ensemble average was made over the
10,000 samples.  Note that by adding the noise to the Monte Carlo samples
one could similarily calculate the total covariance matrix (as we did
for testing).  The advantage of dividing the whole covariance matrix into
three terms is that once we calculate the expression for one value of
$Q^2_{\rm rms-PS}$ we can simply scale it to obtain the results for all
different values of $Q^2_{\rm rms-PS}$ for a given $n$.  The $n$-dependence
of $C^0(\alpha)$ is sufficiently smooth so that we interpolated the matrix
elements of SN and SS evaluated on a grid of values of $n$.  To test the
whole procedure including the values of $N_i$ we have compared the NN term
with the measurement errors for $C(\alpha_i)$ obtained by Smoot et al.
(1992, Fig. 3) and Wright et al. (1992, Fig. 2).  For both data sets our
results agree with the correct values within a few percent.

Given the full covariance matrix it is now straightforward to obtain
maximum-likelihood estimates $\widehat Q_{\rm rms-PS}$ and $\widehat n$
for a given set of data $C(\alpha_k)$.  The covariance matrix of errors
in the parameters may be estimated in the usual way by taking $\Delta\ln
L=0.5$ for one standard deviation.  However, we should not trust these
asymptotic results because of the small numbers of independent data
points given the \cobe\ beam and the intrinsic correlations as well as
our assumption of a normal likelihood function.

There is another reason why our estimator may give biased results.
Our procedure so far does not correctly simulate the data reduction
procedure used by Smoot et al. (1992) and Wright et al. (1992) because
we have not accounted for the fitting and removal of low-order multipole
moments from the maps before the angular correlation function is computed.
If the sky sampling were uniform and complete, this could be accounted
for simply by limiting the range of $l$ used to compute $C^0(\alpha_k)$.
However, incomplete sky coverage couples different multipoles so that the
actual $\langle C(\alpha_k)\rangle$ and covariance matrix differ from what
we give above, resulting in a bias in our maximum-likelihood estimator.

To correct for this and other biases and to determine the variance of our
estimator we resort again to Monte Carlo simulations.  We generate 5000
random sky maps including signal and noise using the estimated values of
$Q_{\rm rms-PS}$ and $n$ for the \cobe\ data as the input parameters.
We fit and remove monopole, dipole, and (optionally) quadrupole
using the correct galactic latitude cut and then compute the angular
correlation function.  These samples are used as the input data in the
maximum-likelihood estimation described above.  The distribution of Monte
Carlo estimates around the true value yields the variance and bias of
the estimate.  We use this bias estimate to correct our results given below.

\section{Results}

Using the Smoot et al. (1992, Fig. 3) data we obtain maximum-likelihood
estimates, before bias correction, of $\widehat n=1.2$ and $\widehat
Q_{\rm rms-PS}=12.2\mk$.  Using the Wright et al. (1992, Fig. 2) data
the corresponding values are $\widehat n=0.9$ and $\widehat Q_{\rm rms-PS}
=13.9\mk$.  As indicated above, our estimator is expected
to be biased.  To compensate for this bias we analyze Monte Carlo
simulations of the data that are made with different choices for the true
$(n,Q_{\rm rms-PS})$ and we try to determine those parameters for which
the mean estimated values equal the ones we obtain from the real data.
We find that the bias in $n$ is less than 0.1, but the bias in
$Q_{\rm rms-PS}$ is significant.  In the Smoot et al. case our bias
for true values $n=1$ and $Q_{\rm rms-PS}=15.0\mk$ is $\Delta\widehat
Q_{\rm rms-PS}=-2.6\mk$ (the mean estimate is $\langle\widehat Q_{\rm
rms-PS}\rangle=12.4\mk$) while in the Wright et al. case it is $\Delta\widehat
Q_{\rm rms-PS}=-2.2$.  The bias is larger for the Smoot et al. analysis
because of the additional quadrupole subtraction applied to the data.
The bias is only weakly dependent on $n$ and $Q_{\rm rms-PS}$.
Thus, the bias-corrected estimates are $(n,Q_{\rm rms-PS})=(1.2,14.8\mk)$
for Smoot et al. and $(0.9,16.1\mk)$ for Wright et al.

In general one is interested in the amplitude of fluctuations for a fixed
value of $n$.  Assuming the scale-invariant slope $n=1$ and combining the
two data sets we obtain $Q_{\rm rms-PS}=15.7\pm 2.6\mk$, where the uncertainty
is taken from our fit to the Wright et al. sample.  There is a strong
anticorrelation between our estimates of $n$ and $Q_{\rm rms-PS}$ and they
cannot be independently determined with high precision.  We find that the
approximate relation between the two parameters is of the form
$$\widehat Q_{\rm rms-PS}=(15.7\pm 2.6)\,\exp[0.46(1-n)]\mk\ .
\eqno(7)$$
Our mean value is slightly higher than that obtained by Adams et al.
(1993) using the $\sigma_{\rm sky}(10^{\circ})$ normalization,
$Q_{\,rms-PS}=15(1\pm 0.2)\exp[0.31(1-n])\mk$.  It is also higher
than that obtained by Scaramella \& Vittorio (1993).  The main reason for
these differences is that the fitting and removal of the dipole and
quadrupole applied to the real data also subtracted some of the higher-order
multipole moments because of nonuniform sky coverage.

Our results for $n=1.0$ agree within the errors with the results
reported by Smoot et al. ($\widehat Q_{\rm rms-PS}=16.7\pm4.7\mk$),
with a slightly lower amplitude and a smaller error bar.  The change
in the amplitude, when combined with the smaller error bar, may not be
enough to significantly improve the consistency of the \cobe\ results
with the low upper limit on smaller angular scales obtained recently
by Gaier et al. (1992).

\acknowledgments
This work was supported by grants NSF AST90-01762 and NASA NAGW-2807.
We thank G. Smoot for providing us with the \cobe\ correlation
function data in binned form, and we acknowledge helpful discussions
with C. Lineweaver and C. Bennett.  We are especially thankful for
the comments and assistance given by the referee, Ned Wright.


\begin{thebibliography}{}
\bibitem{Adams} \reference Adams, F.C., Bond, J.R., Freese, K., Frieman,
    J.A., \& Olinto, A.V. 1993, \prd, 47, 426
\bibitem[Bennett et al. 1992]{bennett92} \reference Bennett, C.L. et al.
    1992, \apjlett, 396, L7
\bibitem[boggess92]{bogges92} \reference Boggess, N. W., et al. 1992,
    \apj, 397, 420
\bibitem{bond} \reference Bond, J.R., \& Efsthathiou, G. 1987, \mnras,
    226, 665
\bibitem{boughn} \reference Boughn, S. P., Cheng, E. S., Cottingham, D. A.,
    \& Fixsen, D. J. 1992, \apjlett, 391, L49
\bibitem{Eadie} \reference  Eadie, W. T. et al. 1971, Statistical Methods
    in Experimental Physics (Amsterdam: North Holland)
\bibitem[Gaier et al. 1992]{gaier92} \reference Gaier, T. et al. 1992,
    \apjlett, 398, L1
\bibitem[gulkis]{gulkis91} \reference Janssen, M. A., \& Gulkis, S. 1992,
    in The Infrared and Submillimetre Sky after \cobe\, ed. M. Signore
    \& C. Dupraz (Dordrecht: Kluwer), 391
\bibitem[Sachs \& Wolfe 1967]{sw67} \reference Sachs, R. K., \& Wolfe,
    A. M. 1967, \apj, 147, 73
\bibitem[Scaramella \& Vittrio 1993]{scara93} \reference Scaramella, R.,
    \& Vittorio, N. 1993, \mnras, 263, L17
\bibitem[Smoot et al. 1990]{smoot90} \reference Smoot, G. F. et al. 1990,
    \apj, 360, 685
\bibitem[Smoot et al. 1992]{smoot92} \reference Smoot, G. F. et al. 1992,
    \apjlett, 396, L1
\bibitem[Torres et al. 1989]{torres89} \reference Torres, S., et al. 1989,
    in Data Analysis in Astronomy III, ed. V. di Gesu, L. Scarsi, \& M. C.
    Maccarone (New York: Plenum), 319
\bibitem[Wright et al. 1992]{wright92} \reference Wright, E. L. et al. 1992,
    \apjlett, 396, L13
\bibitem[Wright et al. 1993]{wright93} \reference Wright, E. L. et al. 1993,
    \cobe\ preprint 93-06, submitted to \apjlett
\end{thebibliography}
\end{document}